\documentclass[twocolumn]{aastex701}
\usepackage{textcomp}

\usepackage{amsmath,amssymb}
\usepackage{amsthm}
\usepackage{gensymb}

\begin{document}

\title{Spherically Polarized Alfv\'en Waves and the Gosling Boost}

\author[0009-0009-0162-2067]{Yuliang Ding}
\affiliation{Department of Earth, Planetary, and Space Sciences, University of California, Los Angeles, CA 90095, USA}
\email{dingyl@ucla.edu}

\author[0000-0002-2381-3106]{Marco Velli}
\affiliation{Department of Earth, Planetary, and Space Sciences, University of California, Los Angeles, CA 90095, USA}
\email{mvelli@ucla.edu}

\author[0000-0001-9570-5975]{Zesen Huang}
\affiliation{Department of Earth, Planetary, and Space Sciences, University of California, Los Angeles, CA 90095, USA}
\email{zesenhuang@g.ucla.edu}

\author[0000-0002-2582-7085]{Chen Shi}
\affiliation{Department of Physics, Auburn University, AL 36849, USA}
\email{chenshi@auburn.edu}

\author[0000-0002-6276-7771]{Lorenzo Matteini}
\affiliation{Imperial College London, London SW7 2AZ, UK}
\email{l.matteini@imperial.ac.uk}

\author[0000-0002-1128-9685]{Nikos Sioulas}
\affiliation{Imperial College London, London SW7 2AZ, UK}
\email{n.sioulas@imperial.ac.uk}

\author[0000-0002-1989-3596]{Stuart D. Bale}
\affiliation{Physics Department, University of California, Berkeley, CA 94720, USA}
\affiliation{Space Sciences Laboratory, University of California, Berkeley, CA 94720-7450, USA}
\affiliation{The Blackett Laboratory, Imperial College London, SW7 2AZ, UK}
\email{bale@berkeley.edu}

\author[0000-0003-2880-6084]{Anna Tenerani}
\affiliation{Department of Physics, The University of Texas at Austin, TX 78712, USA}
\email{Anna.Tenerani@austin.utexas.edu}

\author[0000-0003-2981-0544]{Mingzhe Liu}
\affiliation{Space Sciences Laboratory, University of California, Berkeley, CA 94720-7450, USA}
\email{mingzhe.liu@berkeley.edu}



\begin{abstract}

Alfv\'en waves are thought to play critical roles in solar wind acceleration and plasma heating in the solar corona and inner heliosphere. 
Parker Solar Probe (PSP) has highlighted the role of large amplitude Spherically Polarized Alfv\'en Waves (SPAWs), where the locally constant magnetic field magnitude $|\mathbf{B}|$
together with outward propagation explains the observed one sided radial velocity enhancement - the Gosling boost.
Starting from the MHD equations, we derive the modified wave pressure and Poynting flux under the SPAW condition, and demonstrate both are governed solely by the transverse magnetic fluctuations. 
Using PSP data from Encounters 6--25, we define an unperturbed velocity baseline from the lower 10th-percentile running average and statistically characterize the radial evolution of Alfv\'enic fluctuations. 
The background solar wind velocity shows clear radial acceleration, while the velocity perturbation amplitude $\delta v$ decreases with heliocentric distance. 
This decay is anisotropic between the radial and perpendicular directions, which is a direct consequence of the growing magnetic deflection angle related to the spherical polarization.  
Our results demonstrate that radial velocity enhancements in the young solar wind arise naturally from SPAWs rather than from localized velocity jets, and provide direct observational evidence for the anisotropic radial evolution of SPAWs in the inner heliosphere.

\end{abstract}



\section{Introduction} \label{sec:intro}
Alfv\'enic fluctuations are ubiquitous in the solar wind and are widely regarded as a primary driver of solar wind acceleration and coronal plasma heating \citep{velli_waves_1991, cranmer_self-consistent_2007, verdini_alfven_2007, chandran_alfven_2009,hansteen_solar_2012}. 
Propagating outward through the corona and inner heliosphere, these waves carry significant energy and exert wave pressure on the ambient plasma \citep{tu_mhd_1995, rivera_situ_2024}. 
Understanding their properties and radial evolution is therefore central to the broader problem of solar wind energetics, space plasma physics and prediction of space weather.

\begin{figure*}[ht!]
    \centering
    \includegraphics[scale=0.53]{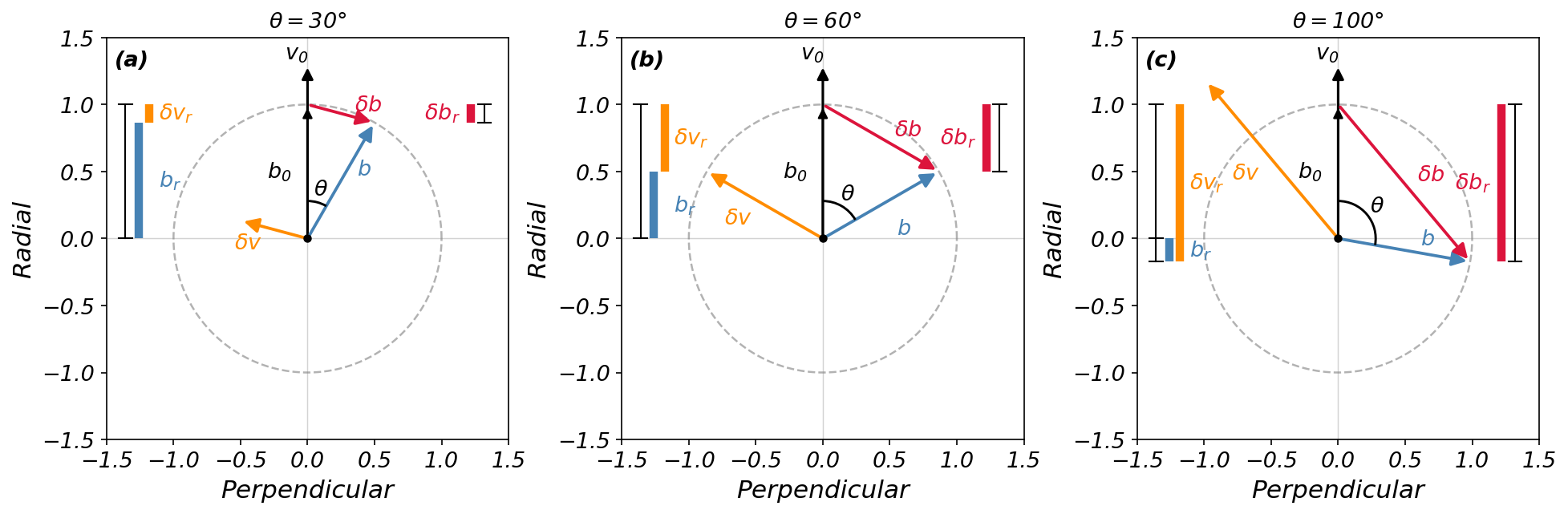}
    \caption{Magnetic and velocity perturbations for Spherically Polarized Alfv\'en Wave, under different deflection angle $\theta$. Magnetic field is normalized to velocity unit by using Alfv\'en speed $b=\frac{B}{\sqrt{\mu_0\rho}}$. The radius of the polarization circle is equal to background magnetic field $b_0$. Stack bar indicates the radial components $\delta v_r$ and $b_r$.}
    \label{fig:sketch}
\end{figure*}

\begin{figure*}[ht]
    \centering
    \includegraphics[width=2.0\columnwidth]{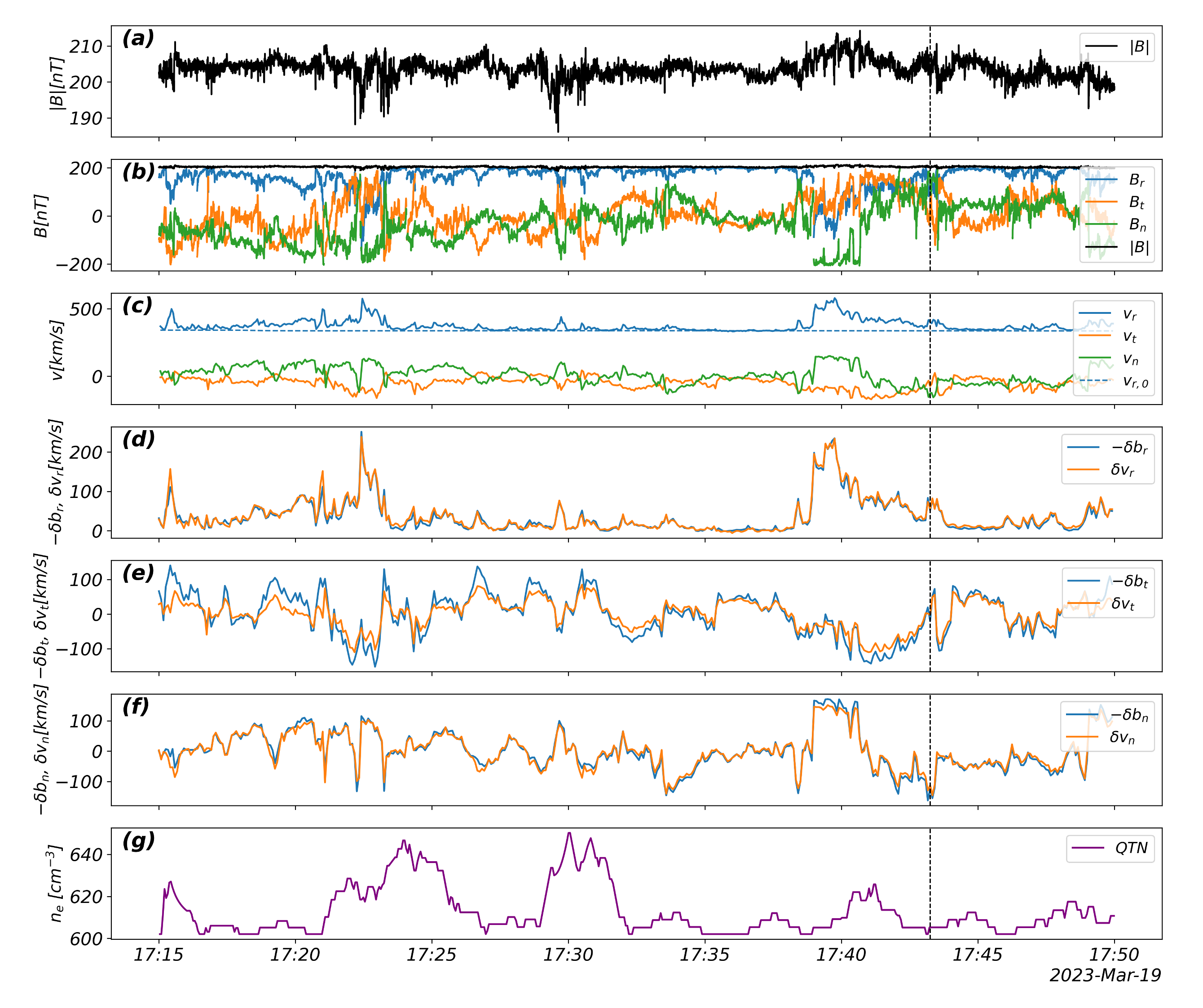}
    \caption{In situ measurements from PSP during Encounter 15. The grey dashed line indicates the selected data point for the left panel of Figure \ref{fig:arrow}.(a)The magnetic field magnitude. (b)Magnetic field magnitude and three components in RTN coordinate. (c)Solar wind velocity components in RTN coordinate. The blue dahed line is the background velocity $v_0$ defined by 10\% lower data average. (d)(e)(f)Three components of magnetic and velocity fluctuations. Magnetic field perturbations are normalized by $\delta b=\frac{\delta B}{\sqrt{\mu_0\rho}}$. (g)The electron density data from quasi-thermal noise (QTN) spectroscopy.}
    \label{fig:exp}
\end{figure*}

Large-amplitude Alfv\'en waves were first observed in the solar wind through plasma and magnetic field measurements from the Mariner 5 spacecraft \citep{belcher_large-amplitude_1971}. The nonlinear MHD theory describing these waves was subsequently developed and progressively refined \citep{barnes_large-amplitude_1974, barnes_nonexistence_1976, goldstein_theory_1974}. 
During the Ulysses mission, arc-polarized Alfv\'en waves were identified and their observational properties systematically characterized \citep{tsurutani_relationship_1994, riley_alfvenic_1995}, with the majority found to be associated with rotational discontinuities \citep{riley_properties_1996}. 
At larger heliocentric distances, observations from the WIND spacecraft confirmed that large-amplitude Alfv\'en waves persist at 1~AU near the Earth \citep{wang_large-amplitude_2012, erofeev_characteristics_2019}.

With Parker Solar Probe (PSP) approaching unprecedentedly close to the Sun, the near-Sun environment was found to be pervaded by large-amplitude magnetic field reversals — the so-called magnetic switchbacks \citep{bale_highly_2019}. 
These structures are predominantly Alfv\'enic, exhibiting strongly correlated or anti-correlated velocity and magnetic field fluctuations alongside a quasi-constant field magnitude $|\mathbf{B}|$, consistent with large-amplitude outward-propagating spherically polarized Alfv\'en waves \citep{woolley_proton_2020, raouafi_parker_2023, badman_properties_2026}.
The origin and evolution of switchbacks have been extensively debated, with numerous simulation studies attempting to reconstruct their structures \citep{toth_theory_2023, tenerani_magnetic_2020, bowen_formation_2025, shoda_turbulent_2021, mallet_exact_2021, shi_analytic_2024}.
Recently, three-dimensional solitary Alfv\'en wave models with quasi-constant $|\mathbf{B}|$ have been established in MHD simulations and shown to exhibit switchback-like structures \citep{huang_solitary_2026}, further highlighting the central role of spherical polarization in the dynamics of the near-Sun solar wind.
A comprehensive understanding of switchbacks therefore requires a thorough characterization of the properties and radial evolution of Spherically Polarized Alfv\'en Waves in the inner heliosphere.

While much of the research has focused on the magnetic characteristics of large-amplitude Alfv\'en waves and switchbacks, the associated velocity signatures have received comparatively less attention. 
\cite{gosling_one-sided_2009} was the first to report that Alfv\'enic fluctuations produce one-sided velocity enhancements along the direction of the background magnetic field.
\citet{matteini_dependence_2014} subsequently extended this framework and provided a theoretical explanation for the dependence of solar wind speed on the local magnetic field orientation.
Later, \citet{matteini_ion_2015} examined the kinetic energy of protons and alpha particles in large-amplitude, highly Alfv\'enic fast solar wind, finding that the motion of protons in SPAWs conserves kinetic energy in the wave frame as a natural consequence of spherical polarization. 
Following these research, \citet{horbury_short_2018} reported short, discrete Alfv\'enic velocity enhancements at 0.3~AU, and discussed their implications for PSP and Solar Orbiter measurements in the inner heliosphere.
Collectively, these studies demonstrate that SPAWs exert a significant influence on the radial solar wind velocity, motivating a systematic investigation of SPAWs and their associated velocity enhancements.
In recognition of \citet{gosling_one-sided_2009} and his pioneering identification of these one-sided velocity enhancements, we propose the term \textit{Gosling Boost} to refer to the one-sided radial velocity enhancements produced by SPAWs throughout this paper.

In this work, we develop a complete theoretical framework for the Gosling Boost produced by the spherical polarization.
Starting from the MHD equations, we derive the modified wave pressure and Poynting flux under the SPAW condition.
Using PSP in situ measurements, we present observational evidence for SPAWs in the inner heliosphere and statistically characterize the radial evolution of the background velocity, velocity fluctuations, and magnetic perturbations.

This paper is organized as follows. 
Section~\ref{sec:theory} exhibits the conditions under which SPAWs produce one-sided radial velocity enhancements and establishes that the associated wave pressure and Poynting flux are determined entirely by the transverse fluctuations. 
Section~\ref{sec:stats} presents a statistical analysis of the radial evolution of Alfv\'enic fluctuations and demonstrates that the observed anisotropy between the radial and perpendicular directions is a natural consequence of spherical polarization. 
Section~\ref{sec:discussion} discusses the implications of these results in the context of existing research. 
Section~\ref{sec:summary} summarizes our conclusions.

\section{Theoretical Model} \label{sec:theory}

\subsection{Spherical Polarization Leads to One-sided Velocity Enhancements}\label{subsec:oneside}
We present a brief analytical proof demonstrating that SPAWs produce one-sided radial velocity enhancements when the background magnetic field and bulk solar wind velocity are both radially aligned. 
The magnetic field and velocity are decomposed as:
\begin{equation}
    \mathbf{B} = B_0\mathbf{\hat{r}} + \mathbf{\delta B_{\parallel}} + \mathbf{\delta B_{\perp}}
    \label{eq:B_decompose}
\end{equation}
\begin{equation}
    \mathbf{v} = v_0\mathbf{\hat{r}} + \mathbf{\delta v_{\parallel}} + \mathbf{\delta v_{\perp}}
    \label{eq:V_decompose}
\end{equation}
where both the background magnetic field $B_0$ and the unperturbed bulk velocity $v_0$ are oriented in the radial direction $\hat{\mathbf{r}}$, and the wave perturbations $\delta\mathbf{B}$ and $\delta\mathbf{v}$ are decomposed into components parallel and perpendicular to this direction. 
Assuming a locally constant field magnitude, $|\mathbf{B}| \simeq |B_0|$, and expanding the square of Equation~\ref{eq:B_decompose}, we obtain:
\begin{equation}
\label{eq:amplitude_relation}
    \delta B_\parallel = -\frac{\delta B^2}{2B_0}
\end{equation}
indicating that $\delta B_\parallel$ is always opposite in sign to $B_0$. For outward-propagating Alfv\'en waves, the velocity and magnetic field fluctuations are related by $\delta\mathbf{v} = -\mathrm{sign}(B_0)\,\delta\mathbf{B}/\sqrt{\mu_0\rho}$, from which $\delta v_\parallel$ follows:
\begin{equation}
    \delta v_\parallel = \mathrm{sign}(B_0)\frac{\delta B^2}{2B_0\sqrt{\mu_0\rho}}>0
\end{equation}
The radial velocity fluctuation is therefore strictly positive. 
As illustrated in Figure~\ref{fig:sketch}, the local magnetic field vector $\mathbf{B}$ traces the surface of a constant-$|\mathbf{B}|$ sphere, and the induced radial velocity perturbation $\delta v_\parallel$ remains exclusively positive. 
This result extends the interpretation of \citet{matteini_dependence_2014} to the special case in which the background magnetic field is closely aligned with the solar wind velocity.

This conclusion holds under the assumptions of radially aligned fields, high Alfv\'enicity, and outward wave propagation — conditions well supported by PSP in situ observations in the inner heliosphere. Figure \ref{fig:exp} shows a 35-minute slice of PSP data from Encounter 15. Panels (a), (b) display high-resolution magnetic field measurements from the FIELDS instrument \citep{bale_fields_2016}, and panel (c) shows the solar wind velocity measured by the Probe ANalyzer for Ions (SPAN-I) \citep{livi_solar_2022}.
The solar wind speed of the selected period is super-Alfv\'enic and the heliocentric distance is 28 solar radii.
In the first two panels, magnetic field magnitude $\mathbf{|B|}$ is nearly constrained throughout the interval.
Several large amplitude magnetic reversals are evident, each accompanied by a corresponding enhancement in the radial velocity.
To separate the unperturbed fields $v_0$ and $B_0$ from Alfv\'enic fluctuations, we apply a 30-minute running average, retaining only the lower 10th percentile for the radial velocity and the upper 10th percentile for the radial magnetic field in each window.
For the the transverse components, a standard 30-minute running mean is used as the background, consistent with the radial alignment assumption.
The perturbations are then obtained by subtracting the corresponding background values from the instantaneous local measurements.
Panels (d), (e), and (f) show the magnetic and velocity perturbations resampled to a 5-second cadence, with magnetic fluctuations normalized to velocity units via $\delta b = \delta B/\sqrt{\mu_0\rho}$. 
The strong agreement between $\delta b$ and $\delta v$ in all three components confirms that the fluctuations are highly Alfv\'enic throughout the interval.
Panel (g) shows the electron number density smoothed with a 5-minute running average, derived from Quasi-Thermal Noise (QTN) spectroscopy \citep{moncuquet_first_2020}, which has served as a critical calibration standard for PSP plasma measurements \citep{liu_total_2023}.
The density fluctuations remain small throughout the interval, with a relative variation below 8\%, consistent with the locally constant density assumption adopted in our theoretical derivations.

\begin{figure}[ht]
    \centering
    \includegraphics[width=1\columnwidth]{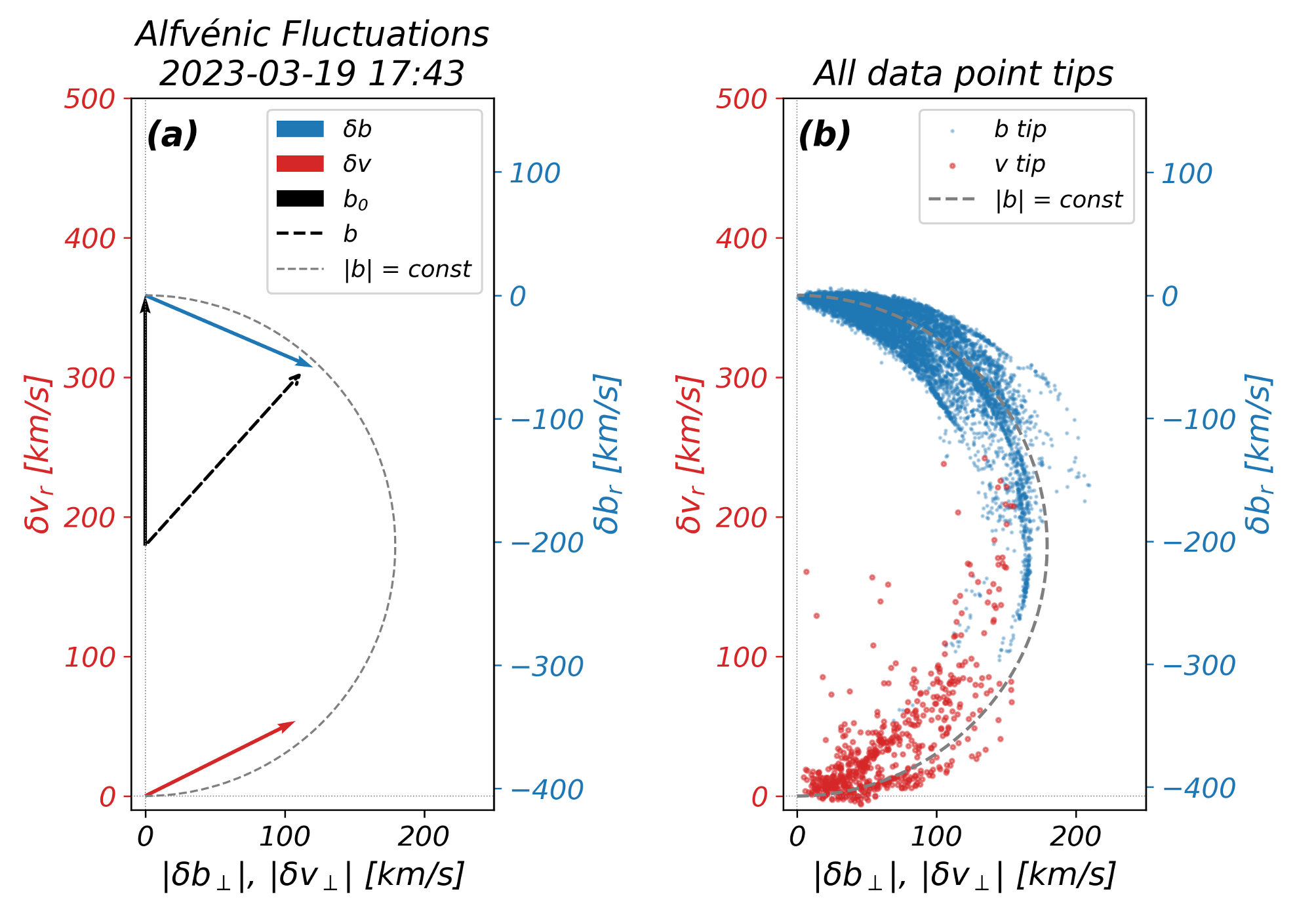}
    \caption{The vector arrow and arrow tips for spherically polarized Alfv\'enic fluctuations. The x-axis is the absolute value. (a)Single data point for a selected time. (b)All data points for the selected stream.}
    \label{fig:arrow}
\end{figure}

To further verify the spherical polarization, we select a representative data point (indicated by the dashed line in Figure \ref{fig:exp}) and display the local magnetic field and Alfv\'enic perturbations in panel (a) of Figure \ref{fig:arrow}. 
The magnetic and velocity perturbation vectors exhibit a strong anti-correlation — consistent with outward Alfv\'enic propagation for positive $b_0$ — and their tips lie on the same constant-$|\mathbf{B}|$ sphere. Note that the horizontal axis shows the magnitude of the perpendicular components $|\delta b_\perp|$ and $|\delta v_\perp|$, so both vectors appear in the positive half-plane.
Panel (b) shows the vector tips of all data points within the selected interval: both $\delta\mathbf{b}$ and $\delta\mathbf{v}$ trace the surface of the constant-$|\mathbf{B}|$ sphere, including large-amplitude reversals exceeding $90^{\circ}$.
Taken together, these results provide direct observational confirmation that SPAW produces one-sided radial velocity enhancements in the young solar wind.

\begin{figure*}[ht]
    \centering
    \includegraphics[width=2\columnwidth]{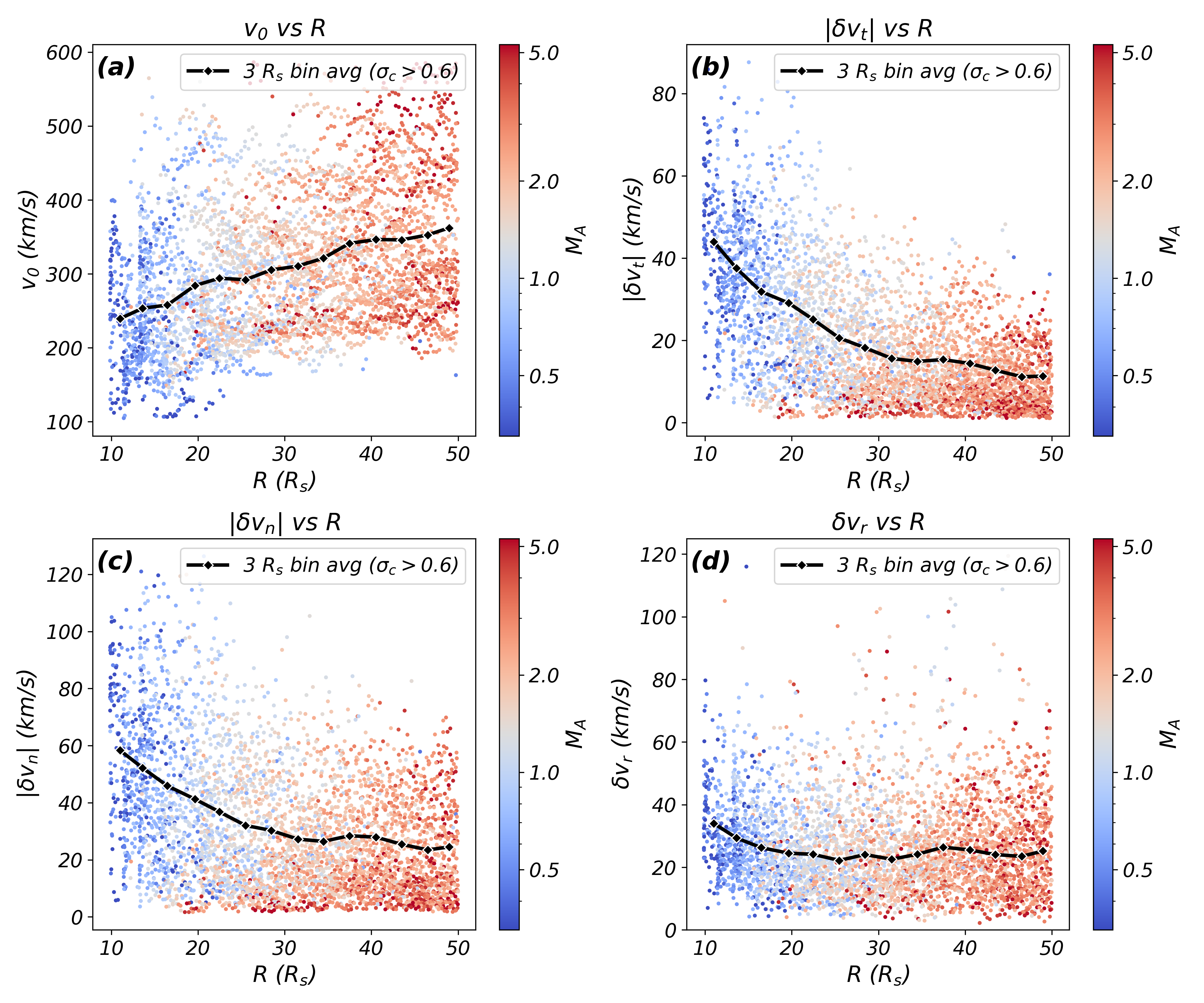}
    \caption{Radial evolution of the unperturbed velocity $v_0$ and velocity fluctuations in RTN coordinate. Dataset includes PSP Encounters 6-25. The scatter points are colored by Alfv\'en Mach number $M_A$. The solid black lines exhibit the averaged data by 3 $R_s$ bins.}
    \label{fig:dV}
\end{figure*}

\subsection{Different Interpretations of SPAWs}\label{subsec:inter}

Figure~\ref{fig:sketch} reveals that the sum $\delta v_r + b_r$ remains constant regardless of the deflection angle $\theta$. 
This can be understood as follows: for a purely transverse Alfv\'en wave propagating along a radial background field, the Alfv\'en speed is $v_{a,tr} = \frac{B_r}{\sqrt{\mu_0\rho}} = b_r$, so the total wave propagation speed in the plasma frame is $v_0 + \delta v_r + v_{a,tr}$. 
The constancy of this sum therefore suggests a natural physical interpretation of SPAWs: they are transverse Alfv\'en waves modified by the constant-$|\mathbf{B}|$ constraint. 
For such a wave, maintaining constant $|\mathbf{B}|$ requires a magnetic perturbation along the radial direction, which reduces the local Alfv\'en speed $v_{a,tr}$. 
A corresponding radial velocity enhancement must therefore arise to preserve the total propagation speed $v_0 + \delta v_r + v_{a,tr}$, producing the one-sided Gosling Boost characteristic of SPAWs.

Additionally, based on previous assumptions for SPAWs, we can derive that both the Alfv\'enic wave pressure and the Poynting flux of SPAWs contain contributions only from the transverse fluctuations.
For outward propagating, spherically polarized Alfv\'en waves, we start from the MHD momentum equation:
\begin{equation}
   \rho (\frac{\partial}{\partial t}+\mathbf{v}\cdot \nabla)\mathbf{v} +\nabla p +\nabla \frac{B^2}{2\mu_0}-\frac{(\mathbf{B}\cdot \nabla)\mathbf{B}}{\mu_0}+\rho \nabla\Phi=0
   \label{momentum1}
\end{equation}

Using equations \ref{eq:B_decompose} and \ref{eq:V_decompose}, assuming the high Alfv\'enicity, and considering $v_r=v_0+\delta v_{\parallel}$, the r-component of equation \ref{momentum1} becomes:
\begin{equation}
    \rho v_r\frac{\partial v_r}{\partial r} +\frac{\partial p}{\partial r}
    + \frac{1}{2\mu_0}\frac{\partial(\delta B^2)}{\partial r}-\frac{1}{\mu_0}\delta B_\parallel\frac{\partial (\delta B_\parallel)}{\partial r}
    +\rho \frac{\partial \Phi}{\partial r}=0
\end{equation}

\begin{equation}
    \rho v_r\frac{\partial v_r}{\partial r} +\frac{\partial p}{\partial r}
    + \frac{\partial}{\partial r}(\frac{\delta B_\perp^2}{2\mu_0})
    +\rho \frac{\partial \Phi}{\partial r}=0\label{end1}
\end{equation}

From equation \ref{end1}, the wave pressure term for purely outward-propagating SPAWs depends exclusively on magnetic perturbations perpendicular to the radial direction.
The full derivation of this equation is elaborated in Appendix \ref{appA}.
Using the same method, we can lead to a similar conclusion on the Poynting Flux of SPAWs, with the full derivation provided in Appendix \ref{appB}.
These results not only yield more accurate estimates of the wave pressure and energy flux in the young solar wind, but also support the interpretation of SPAWs as transverse Alfv\'en waves modified by a constant background magnetic field.
In addition, recent expanding-box simulation observed spherically polarized Alfv\'en wave generated from transverse modes \citep{matteini_alfvenic_2024}, providing evidence for the formation and evolution of SPAWs from traditional transverse waves.

However, it has also been shown that SPAW admits an exact solution to the MHD equations, with Alfv\'en speed $\mathbf{v_{A,\mathrm{sp}}}=\frac{\mathbf{B_0}}{\sqrt{\mu_0\rho}}$ \citep{huang_solar_2024, goldstein_theory_1974}.
In the limiting case of a purely radially aligned solar wind, transverse Alfv\'en waves modified by the constant-$|\mathbf{B}|$ constraint have a different intrinsic Alfv\'en speed, $\mathbf{v_{A,\mathrm{tr}}} = \frac{\mathbf{B}_r}{\sqrt{\mu_0\rho}}$, yet exhibit the same total propagation speed in the solar wind frame. 
These two descriptions are therefore observationally indistinguishable in the inner heliosphere, where the background field is nearly radially aligned. 
At larger heliocentric distances, as the Parker spiral angle increases and the background field departs from radial alignment, the physical characteristics of SPAWs may reveal more information about their relationship to the transverse Alfv\'en modes, offering a potential avenue to distinguish between the two interpretations. 
In this paper, we adopt SPAWs as a unified framework for these perturbations, classifying them by their observational characteristics rather than by their wave mode identity.

\section{Data analysis: Radial evolution of Alfv\'enic fluctuations} \label{sec:stats}

We employ PSP data from Encounters 6–25 for a statistical analysis of the radial evolution of Alfv\'enic perturbations. 
Magnetic data from FIELDS and velocity data from SPAN-I are resampled to a 5-second cadence and merged into a unified dataset. 
The analysis is restricted to 10-50 $R_s$ where both the FIELDS and SPAN-I measurements maintain high reliability.
As a preliminary analysis, Figure~\ref{fig:dV} presents the radial evolution of the background velocity and velocity perturbations.
Using the same method in Section \ref{sec:theory}, we calculate radial unperturbed velocity $v_0$ from the lower 10th percentile of a 30-minute running average, and compute transverse velocity backgrounds from a standard running window average.
The perturbations $\delta v_r$, $\delta v_t$ and $\delta v_n$ are calculated by subtracting the unperturbed values from local measurements.
Data points are colored by the Alfv\'en Mach number $M_A = \frac{|v_0|}{|\mathbf{v}_{A,\mathrm{sp}}|}$, where $\mathbf{v}_{A,\mathrm{sp}} = \frac{\mathbf{B}_0}{\sqrt{\mu_0\rho}}$ is the SPAW Alfv\'en speed. 
In our dataset, 76.7\% of data points have $M_A > 1.0$ and 90.0\% have $M_A > 0.6$, confirming that the majority of the selected intervals are in the super-Alfv\'enic regime.
The Alfv\'enicity of each data point is quantified by the normalized cross helicity, defined as $\sigma_c=\frac{\mathbf{|z_o|}^2-\mathbf{|z_i|}^2}{\mathbf{|z_o|}^2+\mathbf{|z_i|}^2}$, where the Els\"asser variables are $\mathbf{z}_{o,i}=\mathbf{\delta v}\mp sign(B_{r})\frac{\mathbf{\delta B}}{\sqrt{\mu_0\rho}}$\citep{ velli_waves_1991, velli_propagation_1993}. 
The sign of $B_r$ is determined from 30-minute average to exclude the influence of switchbacks. 
In our dataset, 98.4\% of data points have $\sigma_c > 0$ and 81.1\% have $\sigma_c > 0.6$, indicating a strong predominance of outward-propagating Alfv\'enic fluctuations.
In each panel, the radial evolution trend is highlighted by a $3\,R_s$ binned average computed from data points with $\sigma_c > 0.6$, including only the most highly Alfv\'enic intervals.
Panel (a) displays evident acceleration of the base velocity, while panel (b) and (c) show the radial decrease of transverse perturbations.
The radial fluctuation $\delta v_r$ in panel (d), however, remains nearly constant from 20 to 50 solar radii.

\begin{figure*}[ht]
    \centering
    \includegraphics[width=1.5\columnwidth]{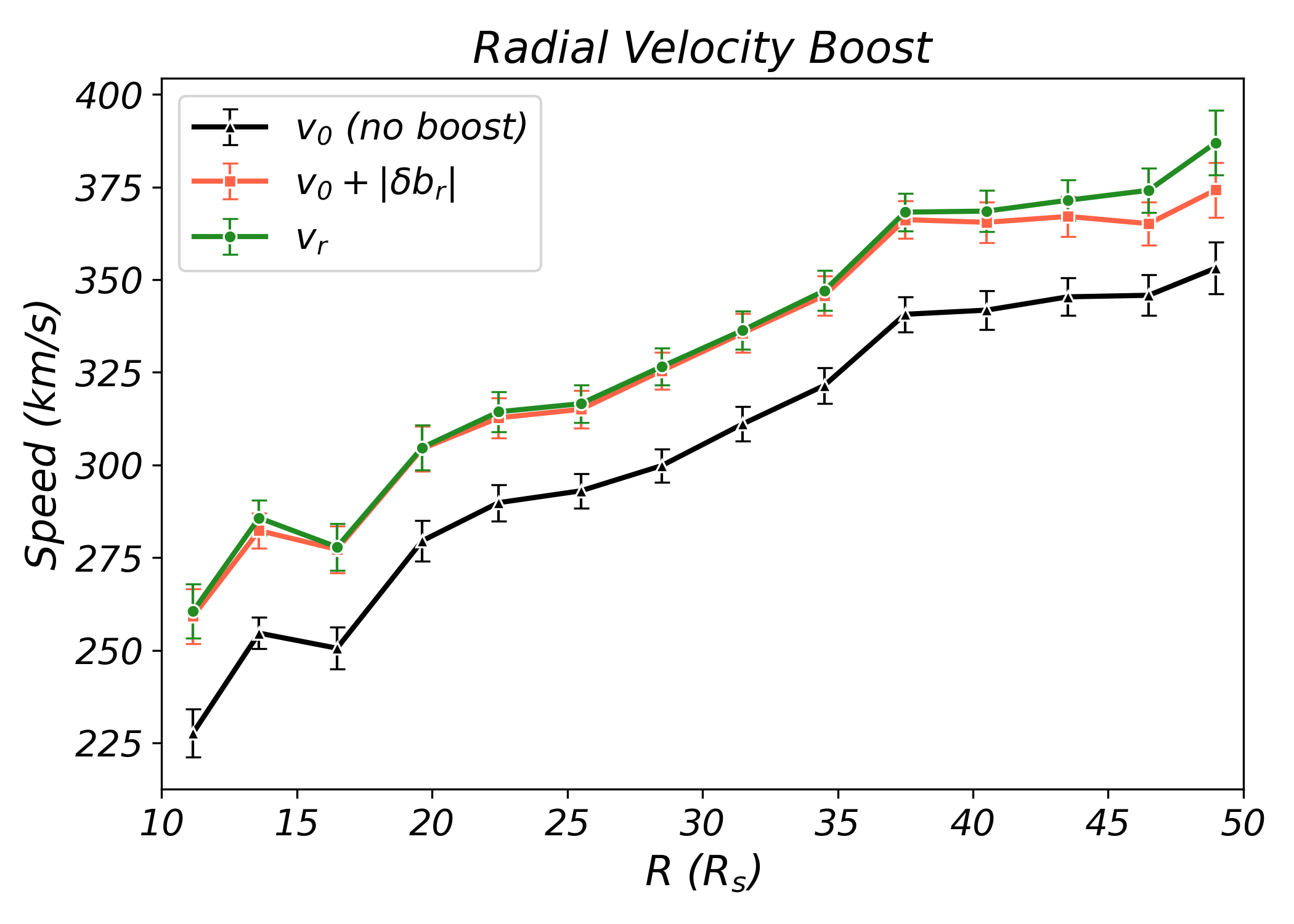}
    \caption{Comparison of the radial velocity baseline with the SPAW-induced velocity boost. The black line shows the unperturbed background velocity $v_0$. The orange line shows $v_0 + \delta b_r$, where $\delta b_r = \delta B_r / \sqrt{\mu_0\rho}$ is the radial magnetic field perturbation normalized to velocity units. The green line shows the measured radial solar wind velocity $v_r$. All quantities are averaged in $3\,R_s$ bins. Error bars denote the standard error of the mean within each bin.}
    \label{fig:v_boost}
\end{figure*}

Figure~\ref{fig:dV} demonstrates that $\delta v_r$, induced by SPAWs, acts as a systematic boost to the background velocity $v_0$. 
To further confirm this, Figure~\ref{fig:v_boost} shows the binned averages of the background velocity $v_0$, the boosted velocity $v_0 + |\delta b_r|$, and the measured radial velocity $v_r$. 
The close agreement between $v_0 + |\delta b_r|$ and $v_r$ confirms both the high Alfv\'enicity of the fluctuations and the SPAW origin of the radial velocity enhancements. 
On average, the solar wind speed is increased by $\sim$25~km/s relative to the background, indicating that the Gosling Boost makes a non-negligible contribution to the radial wind speed.
We note, however, that our dataset is limited to heliocentric distances beyond $\sim$10~$R_s$, leaving the boost amplitude in the lower corona an open question that future modeling and observations closer to the Sun could address.
Beyond $\sim$35~$R_s$, the measured velocity $v_r$ begins to depart from the boosted estimate $v_0 + |\delta b_r|$.
This departure may reflect a combination of physical and instrumental effects: the decreasing Alfv\'enicity and increasing Parker spiral angle at larger heliocentric distances, as well as the reduced reliability of SPAN-I velocity measurements in the outer solar wind.

\begin{figure*}[ht]
    \centering
    \includegraphics[width=2\columnwidth]{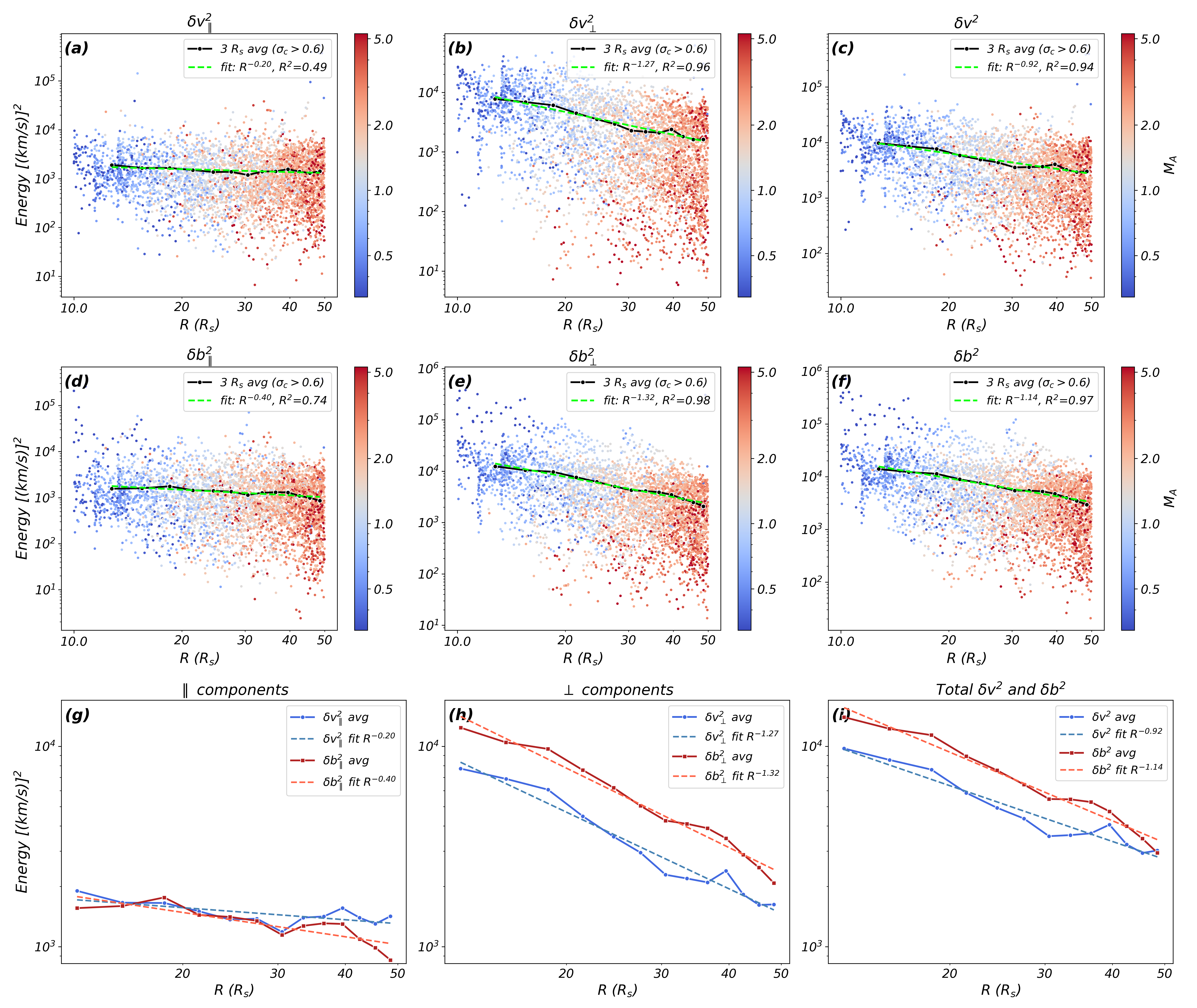}
    \caption{Radial evolution of magnetic and velocity fluctuation energy. The scatter points are colored by Alfv\'en Mach number $M_A$. Panel (a) and (d) for the parallel component, (b) and (e) are for the perpendicular component, (c) and (f) are for the sum of all components. The solid black lines exhibit the averaged data by 3 $R_s$ bins. The green dashed lines are the results from least-squares linear fitting of the averaged data. Panel (g) (h) and (i) show the comparisons of linear fitting results for three columns.}
    \label{fig:dV_dB}
\end{figure*}

To isolate the distinct radial behavior of the parallel and perpendicular components and compare their power-law scalings directly, we compute $\delta v^2$ and $\delta b^2$ for both parallel and perpendicular directions and present their radial profiles in Figure \ref{fig:dV_dB} on a log-log scale, where the parallel component is defined as $\delta v_\parallel^2=\delta v_r^2$ and the perpendicular component as $\delta v_\perp^2=\delta v_t^2 +\delta v_n^2$.
The background magnetic field $B_0$ for the $\delta b_\parallel$ calculation is also determined by the lower or higher 10th-percentile running average, depending on the sign of the averaged radial magnetic field.
To better reveal the underlying radial evolution trends, data points are averaged in 3 $R_s$ bins (solid black lines), and least-squares linear regression is applied to each binned profile (green lines).
The coefficient of determination $R^2$, evaluated in log-log space, quantifies the goodness of fit of the power-law scaling to the binned averages, with values approaching unity indicating a robust fit.

This representation reveals the radial trends more clearly: $\delta v_r$ decreases slowly with heliocentric distance, consistent with the behavior of $\delta b_r$. 
Beyond $\sim$30~$R_s$, however, $\delta v_r$ begins to deviate from $\delta b_r$, as shown in panel~(g), likely owing to a combination of the increasing Parker spiral angle and the reduced reliability of SPAN-I velocity measurements at larger heliocentric distances.
The notably low coefficient $R^2$ in panel (a) is likewise attributable to this deviation, which causes the parallel velocity profile to depart from a single power law across the full radial range.

The transverse fluctuations in Figure \ref{fig:dV_dB} exhibit high Alfv\'enicity, and have steeper slope index compared to the radial direction. 
In panel (h), the energy in magnetic field fluctuation $\delta B_\perp^2$ is larger than $\delta v_\perp^2$, attributable to an instrumental effect for the t components, which also leads to the energy difference in panel (i), the deviation between $\delta b_t$ and $\delta v_t$ in panel (e) of Figure \ref{fig:exp}, and the lower $|\delta v_t|$ compared to $|\delta v_n|$ in Figure \ref{fig:dV}. 

The anisotropy of slope index between radial and perpendicular directions follows the radial evolution of SPAWs. 
For an approximation we adopt the simplified radial density profile $\rho \propto R^{-2}$ \citep{Leblanc_tracing_1998, abraham_radial_2022}, which is consistent with the index of $-1.99$ obtained from our dataset.
From panel (f) we have $\delta b^2 = \frac{\delta B^2}{\mu_0\rho}\propto R^{-1.14}$, giving $\delta B^2 \propto R^{-3.14}$, close to the WKB prediction of -3 scaling for large Alfv\'en Mach number. 
The background magnetic field strength in our dataset follows $B_0 \propto R^{-1.8}$, consistent with previous measurements \citep{bale_highly_2019, huang_solar_2024,silwal_evolution_2025}.
Substituting into Equation \ref{eq:amplitude_relation}, this yields to $\delta B_\parallel = -\frac{\delta B^2}{2B_0}\propto R^{-1.34}$, and consequently $\delta b_\parallel=\frac{\delta B_\parallel}{\sqrt{\mu_0\rho}}\propto R^{-0.34}$. 
This result accounts for the slower decrease in the parallel direction, and agrees well with the observed -0.4 in $\delta b_\parallel ^2$ as shown in panel (g).

At 10 $R_s$, the wave energy in the perpendicular direction is significantly greater than that along the radial direction. 
This difference diminishes with increasing heliocentric distance, owing to the shallower radial scaling of the parallel component, which is a behavior that follows naturally from the 
spherical polarization geometry. 
Specifically, continuing from the previous analysis, pure WKB propagation combined with the $R^{-1.8}$ background field scaling gives $\delta B/B_0 \propto R^{0.3}$, indicating that the magnetic deflection angle grows as the solar wind expands outward \citep{matteini_alfvenic_2024, squire_-situ_2020, mallet_exact_2021,bowen_formation_2025,li_radial_2026}. 
As shown in Figure~\ref{fig:arrow}, a larger deflection angle $\theta$ corresponds to a higher ratio of $\delta B_\parallel$ to $\delta B_\perp$, since a greater fraction of the constant-$|\mathbf{B}|$ sphere projects onto the radial direction. 
The observed radial increase of $\delta b_\parallel^2/\delta b_\perp^2$ and $\delta v_\parallel^2/\delta v_\perp^2$ is therefore a natural and quantitatively consistent consequence of SPAW evolution in an expanding solar wind.

\section{Discussion} \label{sec:discussion}

We have shown analytically that, under the condition of radially aligned magnetic field and solar wind velocity, the constant-$|\mathbf{B}|$ constraint necessarily produces a strictly positive radial velocity enhancement $\delta v_\parallel > 0$. 
This one-sided velocity boost in the young solar wind is therefore a natural geometric consequence of SPAW, rather than evidence for localized velocity jets or reconnection-driven ejecta.
The velocity enhancements do not change the propagation speed of the Alfv\'en waves, but do increase the radial solar wind speed and the kinetic energy.
For large amplitude waves, the increased part $\delta v$ can be comparable to the background velocity $v_0$, indicating that the influence from SPAWs is non-negligible.
An important open question concerns the evolution of the radial velocity enhancement at large heliocentric distances, where the Parker spiral angle becomes significant and the radial alignment assumption breaks down.
A more comprehensive analysis using a broader spacecraft dataset is required to answer these questions.

A key result of this work is that both the wave pressure and the Poynting flux of SPAW are determined by the transverse magnetic fluctuations $\langle\delta B_\perp^2\rangle$.
This has important quantitative implications for estimating the energy budget of the young solar wind.
In particular, analyses that compute wave pressure or energy flux from the total magnetic fluctuation energy will systematically overestimate the contribution from Alfv\'enic waves if the parallel component is included.
However, as shown by the comparison of panels (h) and (i) in Figure~\ref{fig:dV_dB}, the magnitude of this overestimation remains modest across the radial distances considered here. 
Therefore, it does not substantially affect existing results on the radial evolution of solar wind energy budgets derived under the transverse wave assumption \citep{rivera_situ_2024, halekas_quantifying_2023}.
It is worth noting that the one-sided nature of the velocity fluctuations complicates the unambiguous separation of different energy budgets, underscoring the importance of carefully distinguishing bulk kinetic energy from wave energy when interpreting velocity measurements in the inner heliosphere.

Our theoretical analysis builds on \citet{hollweg_transverse_1974}, in which the expression for the spatial variation of the Alfv\'en wave amplitude is derived.
We note that in \citet{hollweg_transverse_1974} the background magnetic field and velocity represent time-averaged quantities, whereas in our derivation they correspond to the defined baseline.
This difference in the definition of background field  changes the amplitude of the parallel perturbations, and consequently changes the estimated energy flux and wave pressure.
We argue, however, that the two derivations are not in conflict, since each result follows self-consistently from its respective initial assumptions.
Moreover, Figure~\ref{fig:dV_dB} shows that the fluctuation energy along the radial direction is significantly smaller than that in the transverse directions, confirming that the time-averaged magnetic field remains a reasonable choice of background for estimations in the inner heliosphere \citep{rivera_situ_2024}.
Nevertheless, the contribution of SPAWs to the radial fluctuations should not be neglected entirely: at larger heliocentric distances, where the magnetic deflection angle grows, this contribution is expected to become more significant and will require a more careful treatment.

Our statistical analysis demonstrates the radial scalings of magnetic and velocity fluctuations differ systematically between the radial and perpendicular directions. 
This provides evidence for the radial evolution of SPAWs and is consistent with previous observations and simulations\citep{tenerani_evolution_2021,matteini_alfvenic_2024,bowen_formation_2025}.
Compared to \cite{bowen_formation_2025}, our results show a more pronounced difference between $\delta B_\parallel$ and $\delta B_\perp$, which we attribute to two factors. 
First, \citet{bowen_formation_2025} analyze data from a single encounter (Encounter 10), whereas our dataset spans Encounters 6--25, providing a broader statistical basis. 
Second, \citet{bowen_formation_2025} determine the background field by projecting each magnetic field vector onto a spherical shell, a method that differs from our percentile-based baseline definition. 
When we apply our analysis to the same single-encounter dataset, we recover less pronounced difference between radial scalings, which is consistent with their results, confirming that the difference in the reported anisotropy is attributable to the larger dataset.

Several studies have shown that large-amplitude Alfv\'en waves can propagate while maintaining spherical polarization out to 1~AU \citep{tsurutani_relationship_1994, gosling_one-sided_2009}, although the fraction of highly Alfv\'enic solar wind intervals decreases substantially with heliocentric distance. 
This suggests that a significant proportion of SPAWs lose their high Alfv\'enicity through dissipation during propagation. 
Besides, at large heliocentric distances, the background magnetic field direction for Alfv\'enic solar winds can largely deviate from the Parker spiral (see \cite{gosling_one-sided_2009}), which is possibly related to the propagation of SPAWs. 
To fully understand these phenomena, future work could extend the present analysis to large-amplitude Alfv\'en waves at and beyond 1~AU, and more broadly characterize the radial evolution of SPAWs from the inner heliosphere to larger heliocentric distances using combined datasets from PSP and Solar Orbiter.

\section{Conclusions and Summary}\label{sec:summary}
Using in situ measurements from PSP Encounters 6--25 combined with analytical derivations from the MHD equations, we have investigated the properties and radial evolution of Spherically Polarized Alfv\'en Waves and the associated Gosling Boost in the inner heliosphere. Our main conclusions are as follows:

\begin{enumerate}

    \item When the background magnetic field and solar wind velocity are radially aligned, SPAWs produce strictly positive radial velocity fluctuations $\delta v_\parallel > 0$. The velocity enhancements associated with magnetic switchbacks are therefore a natural geometric consequence of constant magnetic field magnitude $|\mathbf{B}|$, rather than evidence for localized velocity jets.

    \item Both the wave pressure and the Poynting flux of SPAWs are governed solely by the transverse magnetic fluctuations $\langle\delta B_\perp^2\rangle$, providing a more accurate prescription for estimating the Alfv\'enic energy flux in the young solar wind.

    \item Along the radial direction, the Gosling Boost produces a non-negligible increase in the solar wind velocity of $\sim$25~km/s in the inner heliosphere. In the transverse directions, the velocity perturbation amplitudes $|\delta v_t|$ and $|\delta v_n|$ decrease significantly with heliocentric distance, consistent with the behavior of $\delta b_t$ and $\delta b_n$.

    \item The observed anisotropy between the radial and perpendicular fluctuation scaling is quantitatively consistent with the SPAW framework. 
    The parallel component $\delta b_\parallel$ decays more slowly than the perpendicular component, driven by the growing magnetic deflection angle imposed by the constant $|\mathbf{B}|$ constraint.
    This result is in good agreement with the previous observations\citep{tenerani_evolution_2021} and simulations\citep{matteini_alfvenic_2024}.
\end{enumerate}

Taken together, these results demonstrate that spherical polarization plays a fundamental and quantitatively significant role in shaping the velocity boost, energy distribution and the radial evolution of Alfv\'enic fluctuations in the inner heliosphere. 
These findings advance our understanding of Alfv\'enic turbulence in the interplanetary medium and establish SPAW as a physically essential framework for interpreting large-amplitude fluctuations in the young solar wind.
Several important questions nevertheless remain open. 
The physical mechanism responsible for generating and sustaining spherically polarized perturbations — and for maintaining locally constant $|\mathbf{B}|$ — is not yet fully understood, and warrants further investigation through both simulation and observation. 
It also remains to be determined how SPAWs preserve their structure at larger heliocentric distances, where the magnetic spiral geometry becomes increasingly important. 
Finally, a detailed analysis of the MHD energy conservation equation under the SPAW condition would help to fully characterize the energy partition among the parallel, perpendicular, and background components, providing a more complete picture of Alfv\'enic turbulence and energy transport throughout the interplanetary medium. 
These questions will be the focus of future work.

\begin{acknowledgments}
This work is supported by NASA HTMS 80NSSC20K1275 and NASA FINESST NNH24ZDA001N. Y.D. acknowledges support from NSF SHINE 2229566 and NASA FINESST NNH24ZDA001N.
\end{acknowledgments}

\vspace{5mm}

\software{pyspedas \citep{grimes_space_2022}}

\appendix

\section{Wave pressure of SPAW}
\label{appA}
For outward propagating, spherically polarized Alfv\'en waves, we make following assumptions:
\begin{enumerate}
\item The background magnetic field $B_0$ and velocity $v_0$ are oriented in the radial direction. The local magnetic field and velocity can be expressed as:
\begin{equation}
    \mathbf{B} = B_0\mathbf{\hat{r}} + \mathbf{\delta B_{\parallel}} + \mathbf{\delta B_{\perp}}
    \label{eqap:B_decompose}
\end{equation}
\begin{equation}
    \mathbf{v} = v_0\mathbf{\hat{r}} + \mathbf{\delta v_{\parallel}} + \mathbf{\delta v_{\perp}}
    \label{eqap:V_decompose}
\end{equation}

\item For high Alfv\'enicity, assume $\mathbf{\delta v} = -sign(B_r)\frac{\mathbf{\delta B}}{\sqrt{\mu_0\rho}}$

\item The angle brackets $\langle\rangle$ are used for the time-average of the enclosed quantities. 
By choosing a proper time scale, we should have the transverse perturbations vanishing $\langle\delta B_\perp\rangle=\langle\delta v_\perp\rangle=0$.
Assume we are in a time-averaged stationary system, so $ \frac{\partial}{\partial t}=0$.

\item The magnetic field magnitude $|\mathbf{B}|$ and the density $\rho$ are locally constant.

\end{enumerate}

Using all these assumptions, we start from the MHD momentum equation:
\begin{equation}
   \rho (\frac{\partial}{\partial t}+\mathbf{v}\cdot \nabla)\mathbf{v} +\nabla p +\nabla \frac{B^2}{2\mu_0}-\frac{(\mathbf{B}\cdot \nabla)\mathbf{B}}{\mu_0}+\rho \nabla\Phi=0
   \label{momentum2}
\end{equation}
the r-component can be written as:
\begin{equation}
    \rho [(v_0+\delta v_\parallel)\frac{\partial (v_0+\delta v_\parallel)}{\partial r}-\frac{\delta v_\perp ^2}{r}]+\frac{\partial p}{\partial r}+\frac{1}{2\mu_0}\frac{\partial (B_0^2 +2B_0\delta B_\parallel + \delta B^2)}{\partial r} - \frac{1}{\mu_0} [(B_0+\delta B_\parallel)\frac{\partial (B_0+\delta B_\parallel)}{\partial r}-\frac{\delta B_\perp ^2}{r}]+\rho \frac{\partial \Phi}{\partial r}=0
\end{equation}

By using the high Alfv\'enicity assumption, $\delta v_\perp^2$ and $\delta B_\perp^2$ terms cancel each other. Using $v_r=v_0+\delta v_{\parallel}$ we get:
\begin{equation}
    \rho v_r\frac{\partial v_r}{\partial r} +\frac{\partial p}{\partial r}
    + \frac{1}{2\mu_0}\frac{\partial(\delta B^2)}{\partial r}-\frac{1}{\mu_0}\delta B_\parallel\frac{\partial (\delta B_\parallel)}{\partial r}
    +\rho \frac{\partial \Phi}{\partial r}=0
\end{equation}

\begin{equation}
    \rho v_r\frac{\partial v_r}{\partial r} +\frac{\partial p}{\partial r}
    + \frac{\partial}{\partial r}(\frac{\delta B_\perp^2}{2\mu_0})
    +\rho \frac{\partial \Phi}{\partial r}=0\label{end2}
\end{equation}

The key result of this derivation is that, for purely outward-propagating SPAWs, the wave pressure term depends exclusively on magnetic perturbations perpendicular to the radial direction

\section{Poynting flux of SPAW}\label{appB}
Here we show the MHD Poynting vector of SPAW does not contain the fluctuations along the radial direction. 
We keep all the previous assumptions in \ref{appA}, and set the background velocity to 0, which leads to $\mathbf{v_0}=0$ and $\mathbf{v}=\mathbf{\delta v}$.
The Poynting vector is $-\frac{(\mathbf{\delta v}\times \mathbf{B})\times \mathbf{B}}{\mu _0}$, and with our assumptions we have:
\begin{equation}
    \begin{split}
        -\frac{1}{\mu_0}\langle(\mathbf{\delta v}\times\mathbf{B})\times\mathbf{B}\rangle&= \frac{1}{\mu_0}\langle\mathbf{B}^2\mathbf{\delta v} -(\mathbf{\delta v}\cdot\mathbf{B})\cdot\mathbf{B}\rangle\\
        &= \frac{1}{\mu_0}\langle(\mathbf{B_0}+\mathbf{\delta B})^2\mathbf{\delta v}-[\mathbf{\delta v}\cdot(\mathbf{B_0+\delta B})](\mathbf{B_0+\delta B})\rangle\\
        &= \frac{1}{\mu_0}\langle-(\mathbf{B_0}+\mathbf{\delta B})^2\frac{\mathbf{\delta B}}{\sqrt{\mu_0 \rho}}+[\frac{\mathbf{\delta B}}{\sqrt{\mu_0 \rho}}\cdot(\mathbf{B_0+\delta B})](\mathbf{B_0+\delta B})\rangle\\
    \end{split}
\end{equation}

The average of the transverse component vanishes. Continuously we have:

\begin{equation}
    \begin{split}
        -\frac{1}{\mu_0}\langle(\mathbf{\delta v}\times\mathbf{B})\times\mathbf{B}\rangle&= \frac{1}{\mu_0}\langle\frac{1}{\sqrt{\mu_0\rho}}[-(B_0^2+2B_0\delta B_{\parallel}+\delta B^2)\mathbf{\delta B_{\parallel}}+(\delta B^2+B_0\delta B_{\parallel})(\mathbf{B_0}+\mathbf{\delta  B_{\parallel}})]\rangle\\
        &= \frac{1}{\mu_0}\langle\frac{1}{\sqrt{\mu_0\rho}}(-\delta B_{\parallel}^2\mathbf{B_0} + \delta B^2\mathbf{B_0})\rangle\\
        &= \frac{\langle\delta B_{\perp}^2\rangle}{\mu_0\sqrt{\mu_0\rho}}\mathbf{B_0}\\
        &= \frac{\langle\delta B_{\perp}^2\rangle}{\mu_0}\mathbf{v_a}
    \end{split}
\end{equation}
We have used $\mathbf{v_a}=\frac{\mathbf{B_0}}{\sqrt{\mu_0\rho}}$. For an arbitrary background flow velocity $\mathbf{v_0}$, the result generalizes to $(\mathbf{v_a + v_0})\frac{\langle\delta B_\perp^2\rangle}{\mu_0}$. 
If we only consider the r-component, the brackets in the final result can be removed, so we have $(\mathbf{v_a + v_0})\frac{\delta B_\perp^2}{\mu_0}$. 
Analogous to the Alfv\'enic wave pressure, the Poynting vector of SPAW contains contributions only from the transverse fluctuations. This has important implications for accurately estimating the energy flux of the young solar wind.


\bibliography{g_boost}{}
\bibliographystyle{aasjournal}

\end{document}